# Technical Report 460 / 24-07-2013

Rev. No. 4 / 30-09-2013
University of South-East Europe Lumina, Bucharest, ROMANIA

☒ Public ☐ Confidential

## Cross-Sensor Iris Recognition: LG4000-to-LG2200 Comparison

In coordination of PhD. Nicolaie Popescu-Bodorin,
ACSTL Cross-Sensor Comparison Competition Team 2013:

**Professors:** PhD. Nicolaie Popescu-Bodorin[1] – *IEEE Member*, PhD. Lucian S. Grigore[1] – *IEEE Member*, PhD. Valentina E. Balas[2] – *IEEE Senior Member*.

**Students:** Cristina M. Noaica[3] – *IEEE Student Member*, Ionut Axenie[4], Justinian R. Popa[5], Cristian Munteanu[5], Victor C. Stroescu[6], Ionut A. Manu[7], Alexandru Herea[7], Kartal M. Horasanli[8], Iulia M. Motoc[9] – *IEEE Student Member*.

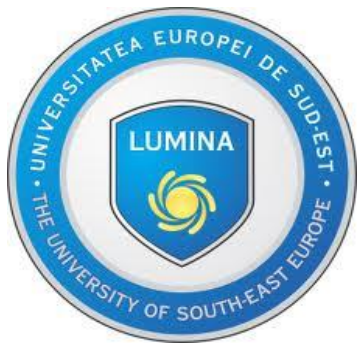

Corresponding author: PhD. Nicolaie Popescu-Bodorin ( bodorin # ieee . org )

**Applied Computer Science Testing Laboratory (ACSTL),
Lumina Multidisciplinary Research Excellence Center (LMREC),
University of South-East Europe Lumina (USEEL), Bucharest, ROMANIA.**

**June-September, 2013**

[1] Applied Computer Science Testing Laboratory, LMREC, University of South-East Europe Lumina, Bucharest, RO;
[2] ACSTL Team Member, Associate Professor, Faculty of Engineering, 'Aurel Vlaicu' University of Arad, Arad, RO;
[3] ACSTL Volunteer & Student Team Member, 2nd year MSc Student in Artificial Intelligence, Comp. Sci. Dept., University of Bucharest, RO;
[4] ACSTL Volunteer & Student Team Member, 3rd year Student in Telecommunications, University of S-E Europe Lumina, Bucharest, RO;
[5] ACSTL Volunteer & Student Team Member, 4th year HS Student in Computer Science, 'A.I. Cuza' HS, Bucharest, RO;
[6] ACSTL Volunteer & Student Team Member (LMREC, USEEL), 2nd year HS Student at 'Goethe' Deusche National College, Bucharest, RO;
[7] ACSTL Volunteer & Student Team Member, 4th year HS Student in Maths & Comp. Sci., 'Tudor Vianu' National College, Bucharest, RO;
[8] ACSTL Volunteer & Student Team Member, 1st year MSc Student in Communication Management, 'Politehnica' University of Bucharest, RO;
[9] ACSTL Volunteer & Student Team Member (LMREC, USEEL), 2nd year MSc Student in Information Security & Biometrics, School of Engineering and Digital Arts, University of Kent, Canterbury, Kent, UK;

# Foreword

This Technical Report[10] describes the results obtained by *ACSTL Cross-Sensor Comparison Competition Team 2013*[11] (Applied Computer Science Testing Lab[12], Lumina Multidisciplinary Research Excellence Center, University of South-East Europe Lumina, Bucharest, RO) in coordination of N. Popescu-Bodorin[13], during the *Cross-Sensor Comparison Competition 2013*[14] organized by Amanda Sgroi[15], Kevin Bowyer[16] and Patrick Flynn[17] (Computer Vision & Research Lab[18], Department of Computer Science & Engineering, Notre Dame University, Notre Dame, IN 46556, USA) within the framework of IEEE-BTAS-2013 Conference.

# Abstract

Cross-sensor comparison experimental results reported here show that the procedure defined and simulated during the *Cross-Sensor Comparison Competition 2013* by our team for migrating / upgrading LG2200 based to LG4000 based biometric systems leads to better LG4000-to-LG2200 cross-sensor iris recognition results than previously reported, both in terms of user comfort and in terms of system safety. On the other hand, LG2200-to-LG400 migration/upgrade procedure defined and implemented by us is applicable to solve interoperability issues between LG2200 based and LG4000 based systems, but also to other pairs of systems having the same shift in the quality of acquired images.

---

[10] http://lmrec.org/acstl/tr-460.pdf
[11] http://lmrec.org/acstl/cross-sensor-competition-team-2013/
[12] http://lmrec.org/acstl/
[13] http://lmrec.org/bodorin/
[14] http://www3.nd.edu/~asgroi/Competition/CrossSensorCompetition.htm
[15] http://engineering.nd.edu/profiles/asgroi
[16] http://www3.nd.edu/~kwb/
[17] http://www3.nd.edu/~flynn/
[18] http://www3.nd.edu/~cvrl/

## I. Iris Segmentation, Encoding and Matching Procedures for LG4000-to-LG2200 Cross-Sensor Iris Recognition

The methods used by us to process the data sets given in the *Cross-Sensor Comparison Competition 2013* comply to the ISO/IEC 19794-6 [5] data interchange format standard (Fig. 1). The eye images of KIND 1 are processed in order to detect the pupil center and the pupil boundary. Using this information, a region of interest is selected from the initial eye image and represented as an image of KIND 48, which is further unwrapped as a KIND 16 unsegmented pupil-centered polar image (Fig. 2). The limbic boundary is found in the unwrapped version of the region of interest (KIND 16) by applying CFIS2 (Circular Fuzzy Iris Segmentation, second variant, [1]). As a result, the unwrapped iris segment is obtained (Fig. 3). The segmented iris image (Fig. 3) is further encoded as a binary iris code (Fig. 4) using the Log-Gabor encoder, as in [1].

Hamming similarity (HS) expresses the matching between two iris codes:

$$HS = \frac{\|codeA == codeB\|}{Length(codeA)} \qquad (1)$$

In the scenario of migrating/upgrading a LG2200 based system to a LG4000 based system, we consider that the LG2200 iris codes are simultaneously available for comparison with LG4000 iris code candidates. This allows us to assemble LG2200 binary iris codes representing the same eye as an LG2200 digital identity [2] and to compute – for any LG4000 iris code - a fuzzy degree of membership to these LG2200 digital identities. Segmentation, digital identity encoding and matching procedures used for obtaining the results presented here are all proprietary. LG2200-toLG400 migration/upgrade procedure defined and implemented by us is applicable to solve interoperability issues between LG2200 based and LG4000 based systems, but also to other pairs of systems having the same shift in the quality of acquired images.

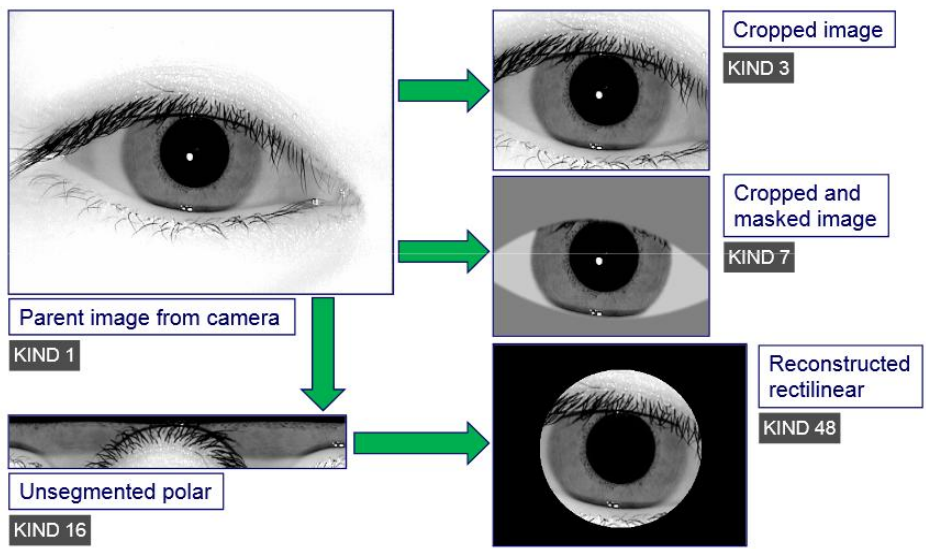

Fig. 1: ISO/IEC 19794-6 data interchange format standard

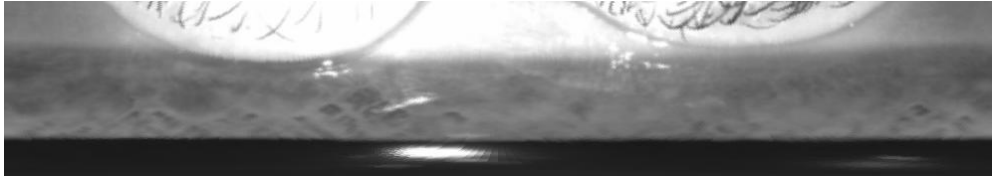
Fig. 2: Unwrapped region of interest (KIND 16)

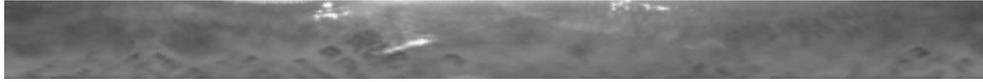
Fig. 3: Segmented polar image

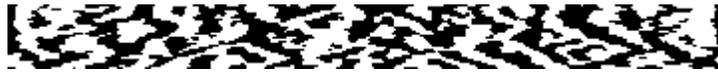
Fig. 4. Iris code of size 32x360

The following sections present the results obtained for LG4000 and LG2200 small, medium and large datasets, respectively. Each section contains figures illustrating the imposter / genuine score distributions and FAR / FRR curves, in linear and logarithmic scales.

The figures 5, 6, 10, 11, 15, 16 and Table 1 illustrate the separation of the genuine and imposter score distributions (obtained for small, medium and large databases) in terms of their means and their standard deviations.

| Comparison | Distribution | Mean | Standard Deviation | Decidability |
|---|---|---|---|---|
| Small LG4000-to-LG2200 cross-sensor recognition: | Imposter | 0.49845 | 0.0136 | 5.5779 |
| | Genuine | 0.66322 | 0.0395 | |
| Medium LG4000-to-LG2200 cross-sensor recognition: | Imposter | 0.49874 | 0.0127 | 5.6386 |
| | Genuine | 0.66038 | 0.0385 | |
| Large LG4000-to-LG2200 cross-sensor recognition: | Imposter | 0.49882 | 0.0124 | 5.7255 |
| | Genuine | 0.66065 | 0.0380 | |

**Table 1: Means and Standard Deviations**

For each comparison test (small, medium and large), a decidability index is computed accordingly to the formula (2),

$$d' = \frac{|\mu_1 - \mu_2|}{\sqrt{\frac{1}{2}(\sigma_1^2 + \sigma_2^2)}} \qquad (2)$$

where:
- $d'$ is the decidability index;
- $\mu_1$ is the mean of the imposter scores;
- $\mu_2$ is the mean of the genuine scores;
- $\sigma_1$ is the standard deviation of the imposter scores;
- $\sigma_2$ is the standard deviation of the genuine scores;

The separation between the genuine and imposter score distributions is also illustrated in Table 2 and figures 6, 11, 16 in terms of minimum genuine similarity score and maximum imposter similarity score. The overlapping between the imposter and genuine score distributions can be viewed as a performance criterion of the recognition technique, but also as a quality measure for the database of eye images. It must be noted that in Table 2 and also in the entire present report, is the SigSets 2013 specification that defines the imposter and the genuine comparisons, not otherwise. However, as the Figures 6, 11, and 16 show, there are some inconsistencies in the way in which the SigSets 2013 specification defines imposter and genuine comparisons: there are indeed some eye images enrolled under wrong IDs - our visual inspection of the databases confirmed that. These cases lead up to the fact that some genuine and some imposter comparison defined by the SigSets 2013 are, in fact, mislabeled. Therefore, some so-called imposter scores defined by the SigSets are actually true genuine scores and some so-called genuine scores defined by the SigSets are actually true imposter scores indeed. Inevitably, this situation alters the FAR and FRR curves, and also the EER points as thresholds (abscises) and values (ordinates) making them to be slightly pessimistic estimates of the actual ones.

The existence of the mislabeled eye images within the *Cross-Sensor Comparison Database* (and consequently the existence of the corresponding mislabeled imposter and genuine comparisons within the SigSets specification) is publicly affirmed here for the first time, despite the fact that the databases are in use for quite a while now. The detection of these cases of mislabeled eye images validates our work and our iris recognition technique. Naturally, a good iris recognition technique must be able to detect mislabeled eye images, especially when the number of eye images within databases is large – which is the case here, in *Cross-Sensor Comparison Competition 2013*.

A clean-up of the *Cross-Sensor Comparison Database* and the corresponding update of the SigSets 2013 specification are both necessary. We will approach this task in our future work.

| Database | Minimum genuine score | Maximum imposter score |
|---|---|---|
| Small | 0.4605 | 0.6747 |
| Medium | 0.4485 | 0.6797 |
| Large | 0.4715 | 0.6947 |

Table 2: Minimum genuine scores and maximum imposter scores

| Database | Figure | FAR, FRR, threshold |
|---|---|---|
| Small | Fig. 9 | (1E-3, 0.009267, 0.5506), (1E-4, 0.01655, 0.5636), (1E-5, 0.02863, 0.5776), (1E-6, 0.1021, 0.6117) |
| Medium | Fig. 14 | (1E-3, 0.00941, 0.5476), (1E-4, 0.01512, 0.5606), (1E-5, 0.02377, 0.5736), (1E-6, 0.06829, 0.6006) |
| Large | Fig. 19 | (1E-3, 0.009723, 0.5476), (1E-4, 0.0144, 0.5596), (1E-5, 0.02361, 0.5726), (1E-6, 0.138, 0.6207) |

Table 3: Triplets of FAR, FRR, and threshold values detected for small, medium and large LG4000-to-LG2200 comparisons

| Database | Figures | EER value @ EER threshold |
|---|---|---|
| Small | Fig. 7, Fig. 8 | 0.005669 @ 0.5386 |
| Medium | Fig. 12, Fig. 13 | 0.006733 @ 0.5356 |
| Large | Fig. 17, Fig. 18 | 0.006167 @ 0.5356 |

Table 4: EER values detected for small, medium and large LG4000-to-LG2200 comparisons

Based on the FAR/FRR curves presented in the figures 7, 8, 12, 13, 17 and 18, the EER points were located at the following thresholds: 0.5386 (Fig. 8), 0.5356 (Fig. 13), and 0.5356 (Fig. 18), with their corresponding values of approximately 5.66E -3, 6.73E -3, and 6.16E -3. For convenience, these data are also collected in Table 4.

Figures 9, 14 and 19 show several triplets (FAR, FRR, threshold) describing four ways of balancing system security (expressed as FAR) and user discomfort (expressed as FRR). These data are also collected in Table 3.

## II. LG4000-TO-LG2200 IRIS RECOGNITION RESULTS OBTAINED
## ON SMALL LG4000 AND LG2200 DATABASES

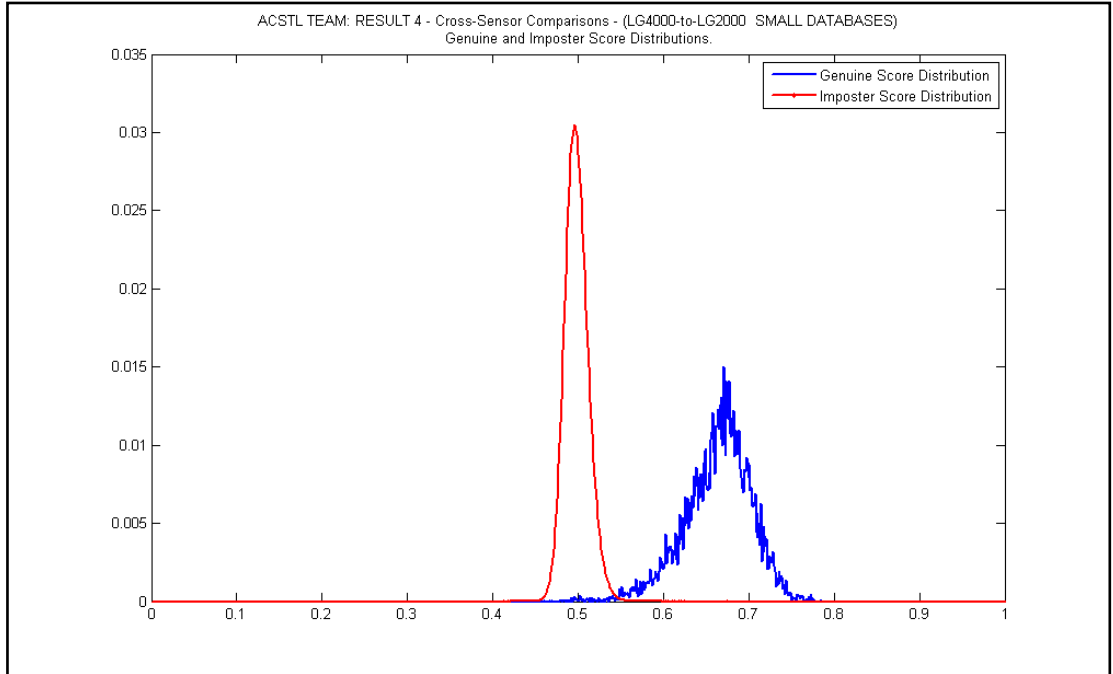

Fig.5: Imposter (μ=0.49845 σ=0.0136) and Genuine (μ=0.66322 σ=0.0395)
Score Distributions (LinearY Scale)

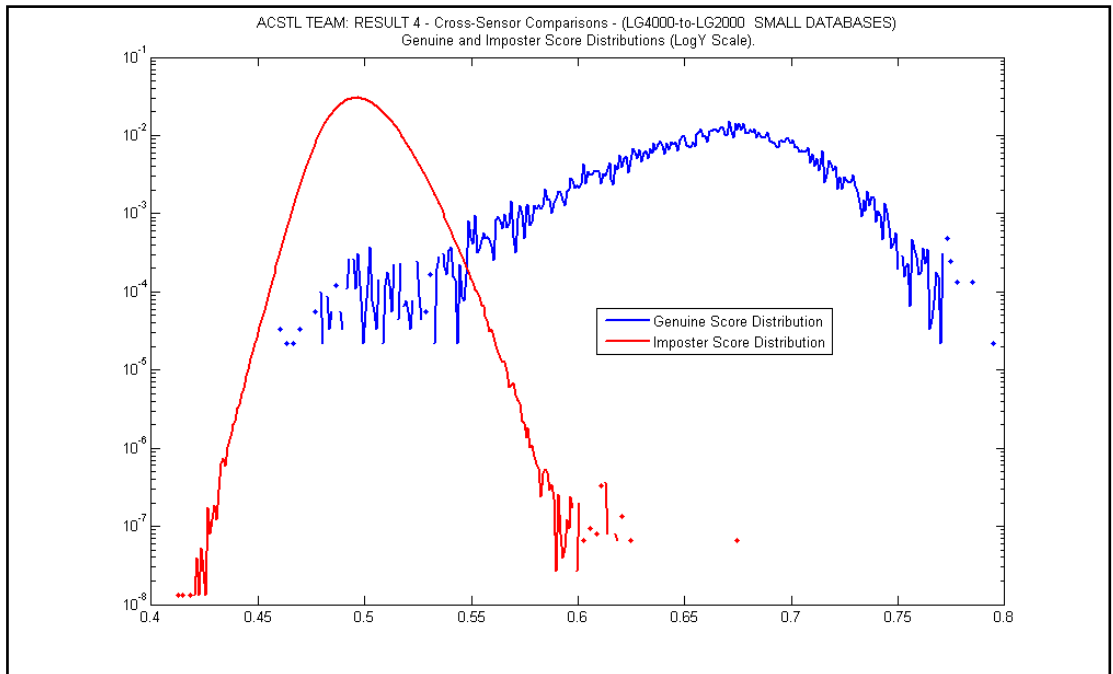

Fig.6 Imposter (μ=0.49845 σ=0.0136) and Genuine (μ=0.66322 σ=0.0395)
Score Distributions (LogY Scale)

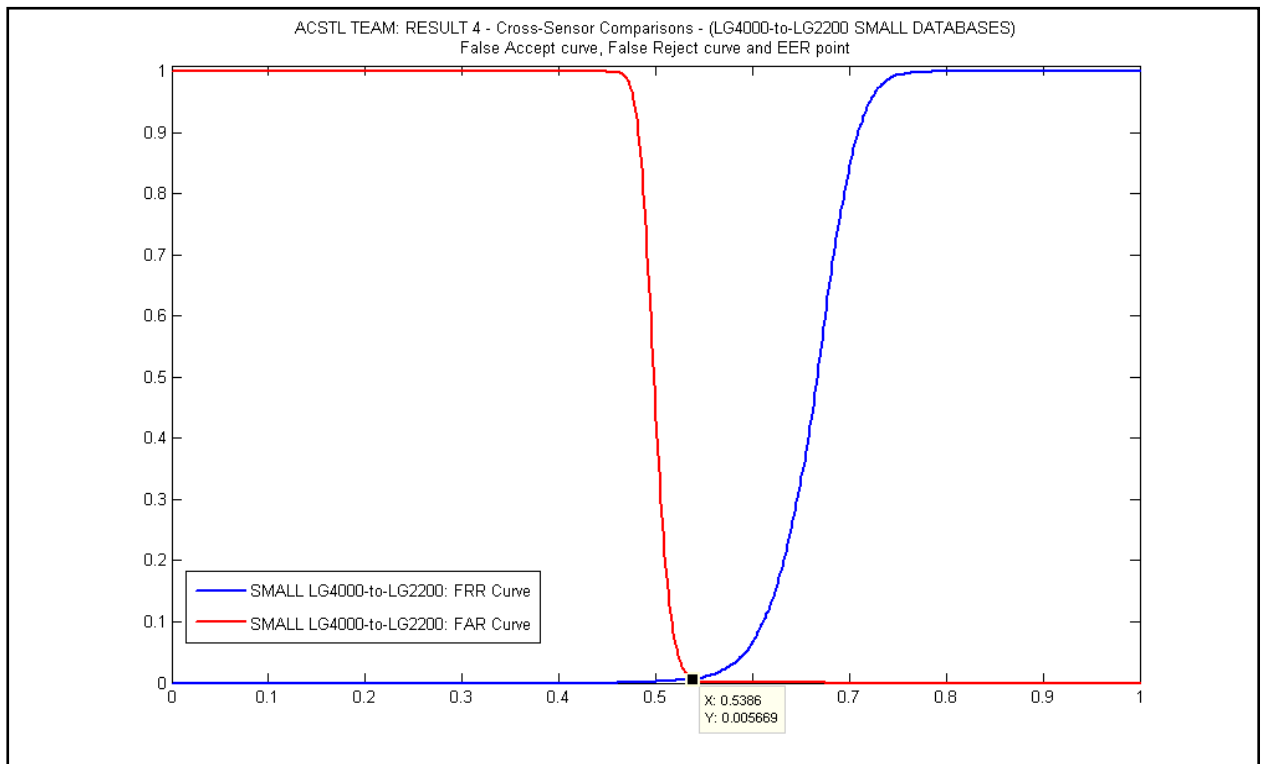

Fig.7: False Accept curve, False Reject curve and EER point (LinearY Scale)

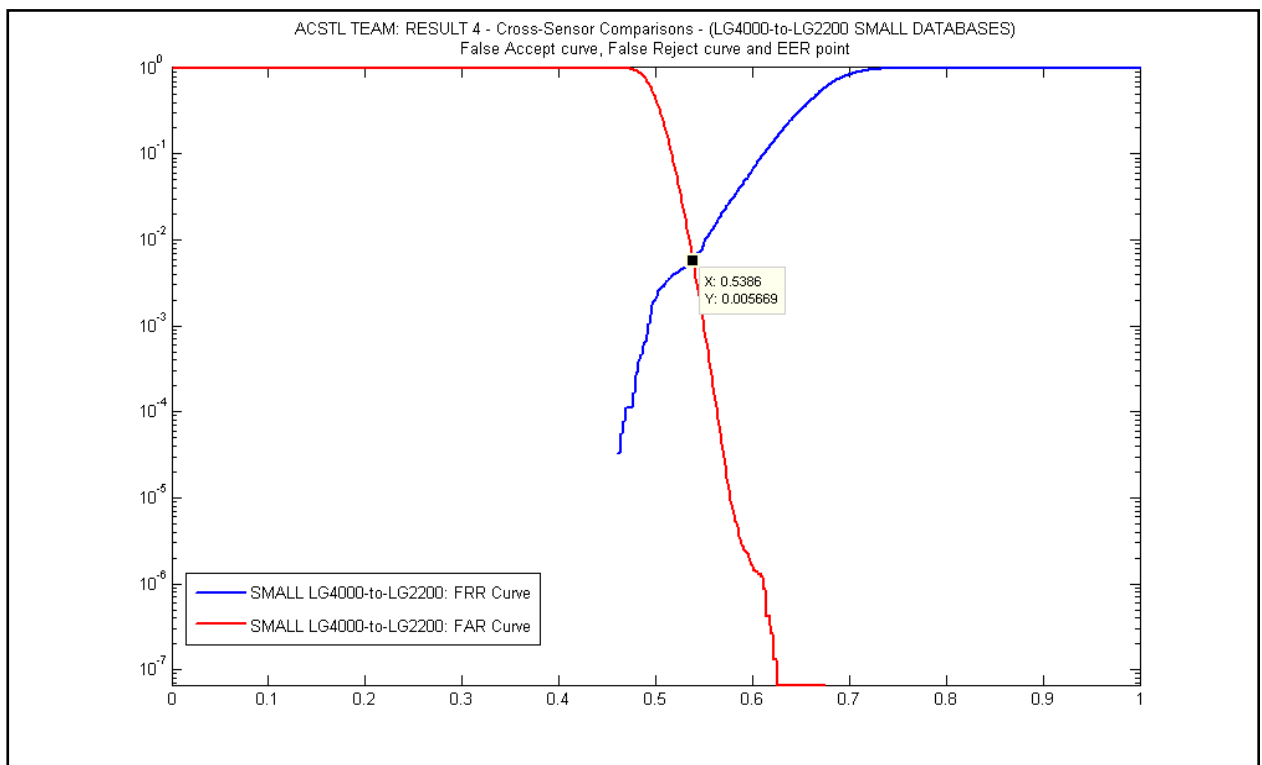

Fig.8: False Accept curve, False Reject curve, and EER point (LogY Scale)

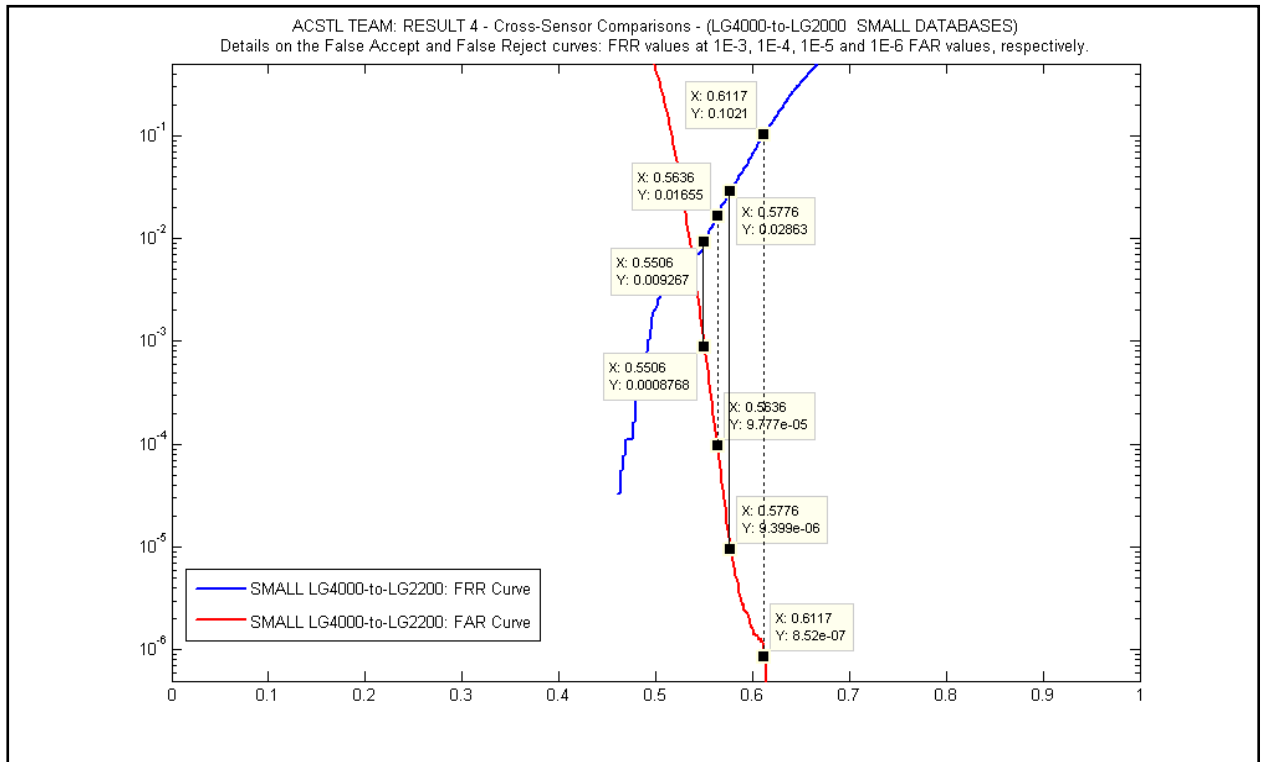

Fig.9: Details on the False Accept curve and False Reject curve (LogY Scale)

# III. LG4000-TO-LG2200 IRIS RECOGNITION RESULTS OBTAINED ON MEDIUM LG4000 AND LG2200 DATABASES

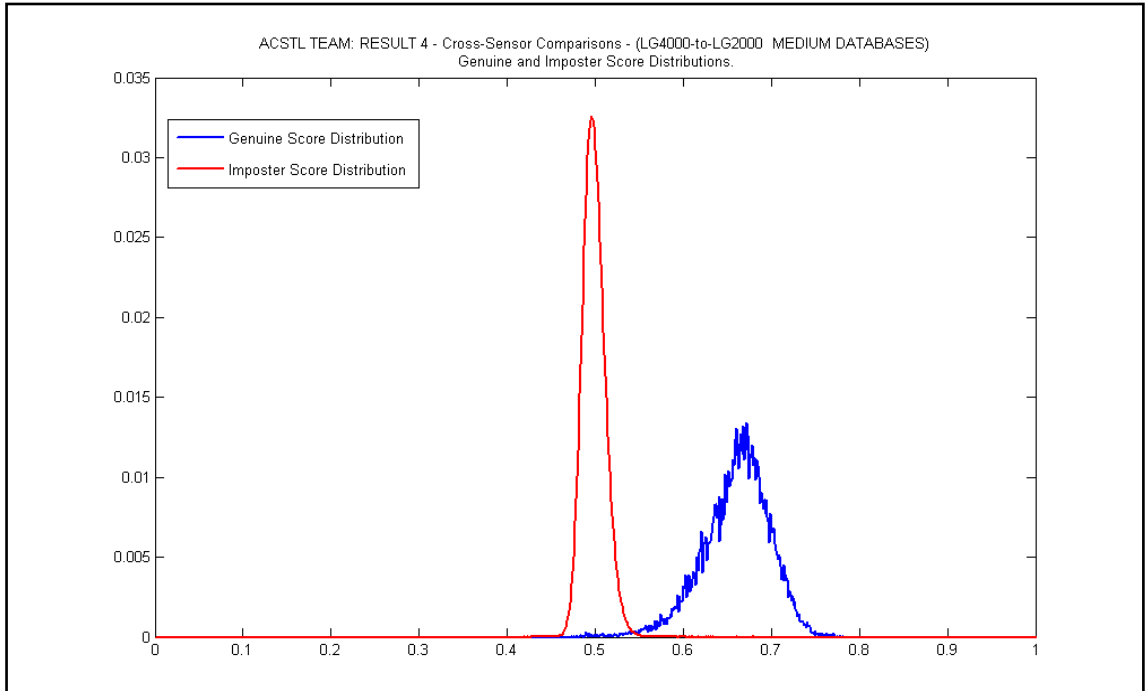

Fig.10: Imposter (µ=0.49874 σ=0.0127) and Genuine (µ=0.66038 σ=0.0385) Score Distributions (LinearY Scale)

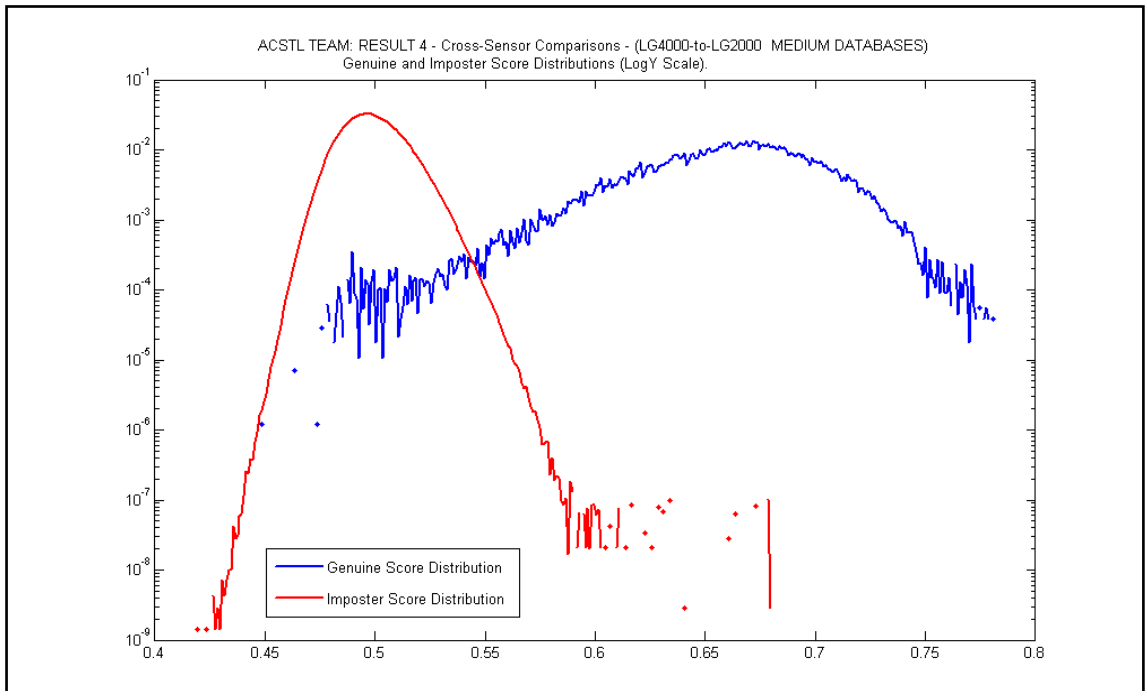

Fig.11: Imposter (µ=0.49874 σ=0.0127) and Genuine (µ=0.66038 σ=0.0385)

Score Distributions (LogY Scale)

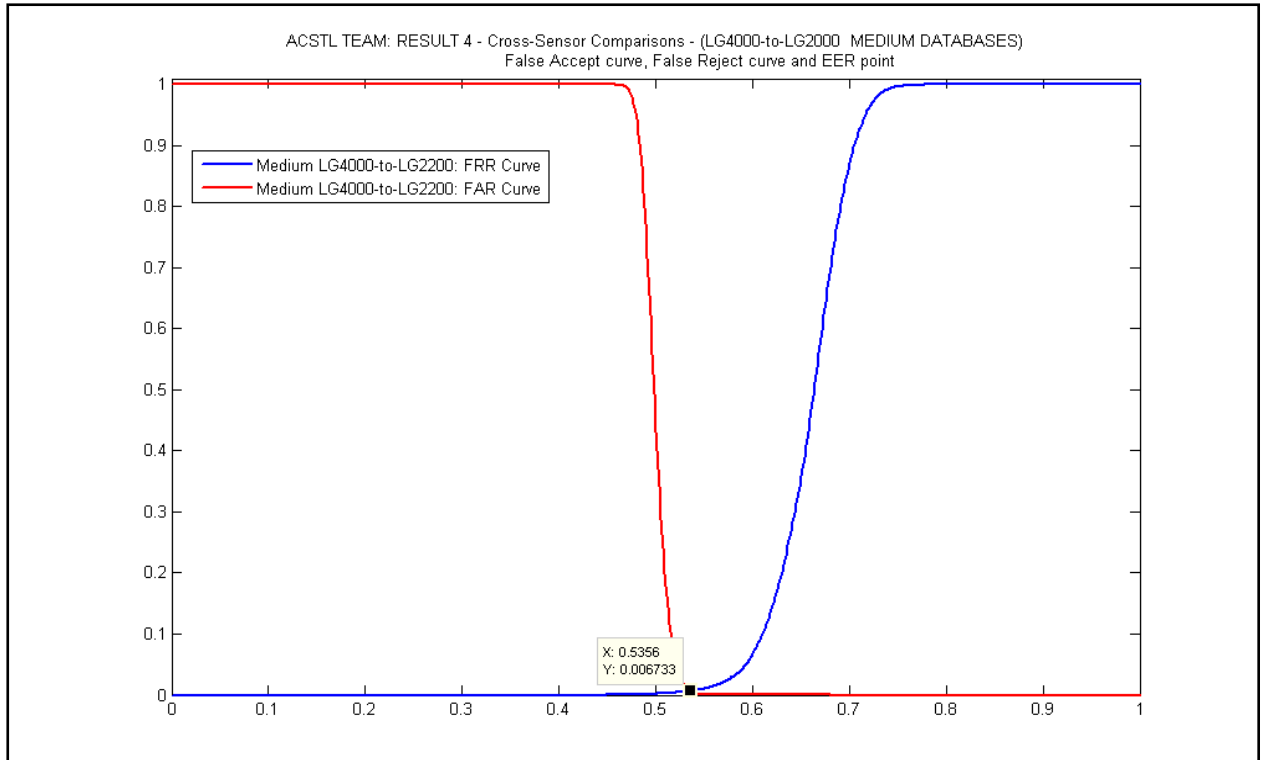

Fig.12: False Accept curve, False Reject curve, and EER point (LinearY Scale)

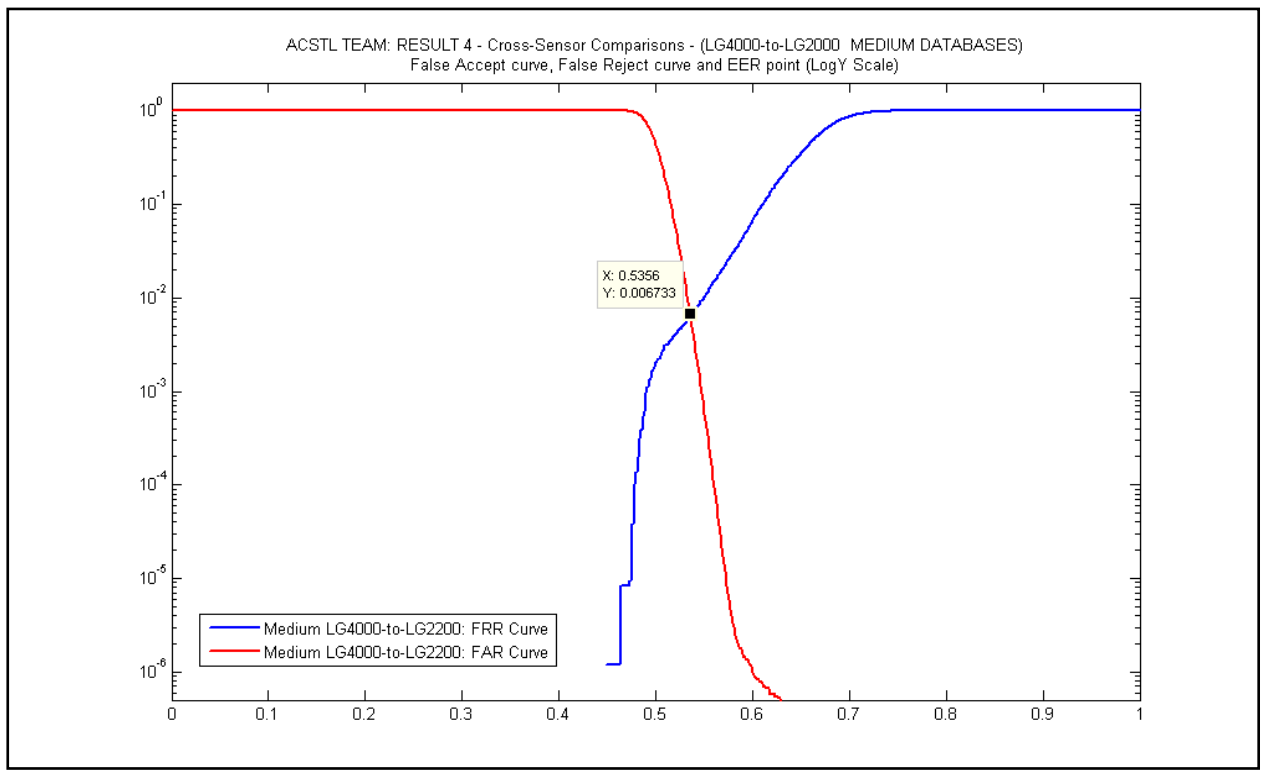

Fig.13: False Accept curve, False Reject curve, and EER point (LogY Scale)

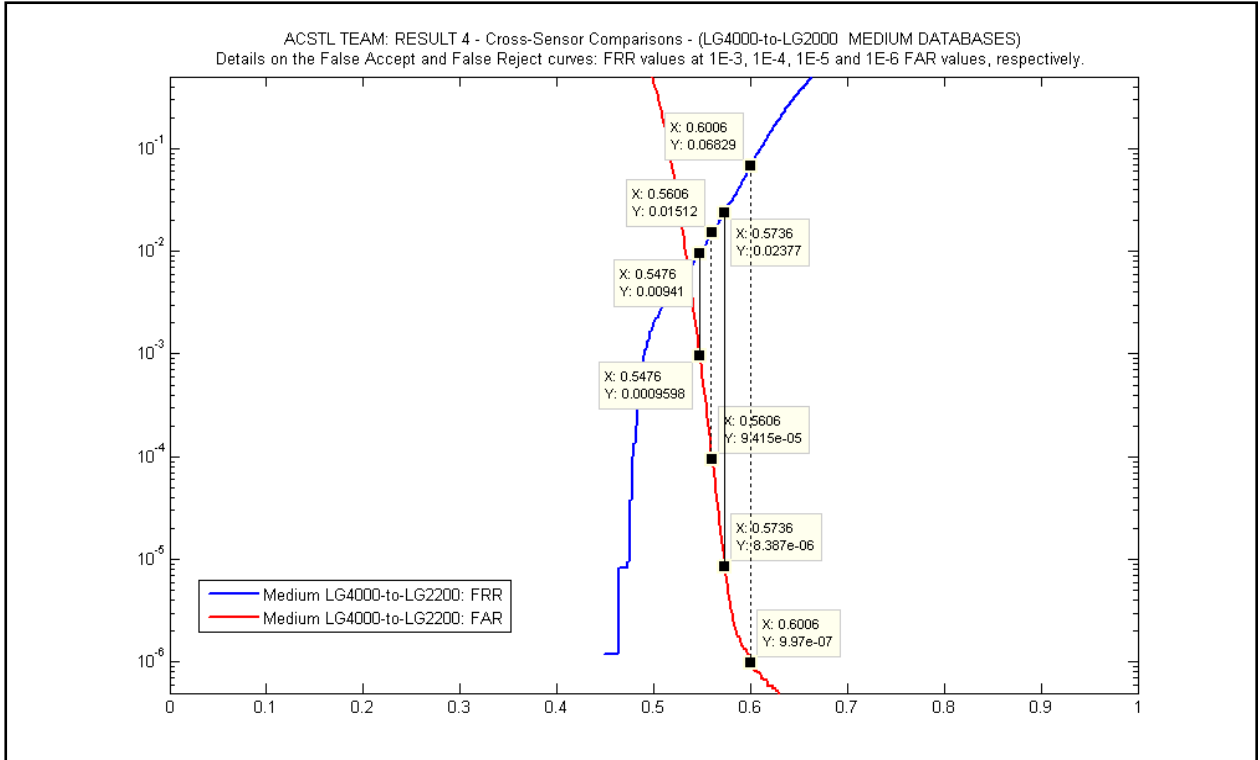

Fig.14: Details on the False Accept curve and False Reject curve (LogY Scale)

## IV. LG4000-TO-LG2200 IRIS RECOGNITION RESULTS OBTAINED ON LARGE LG4000 AND LG2200 DATABASES

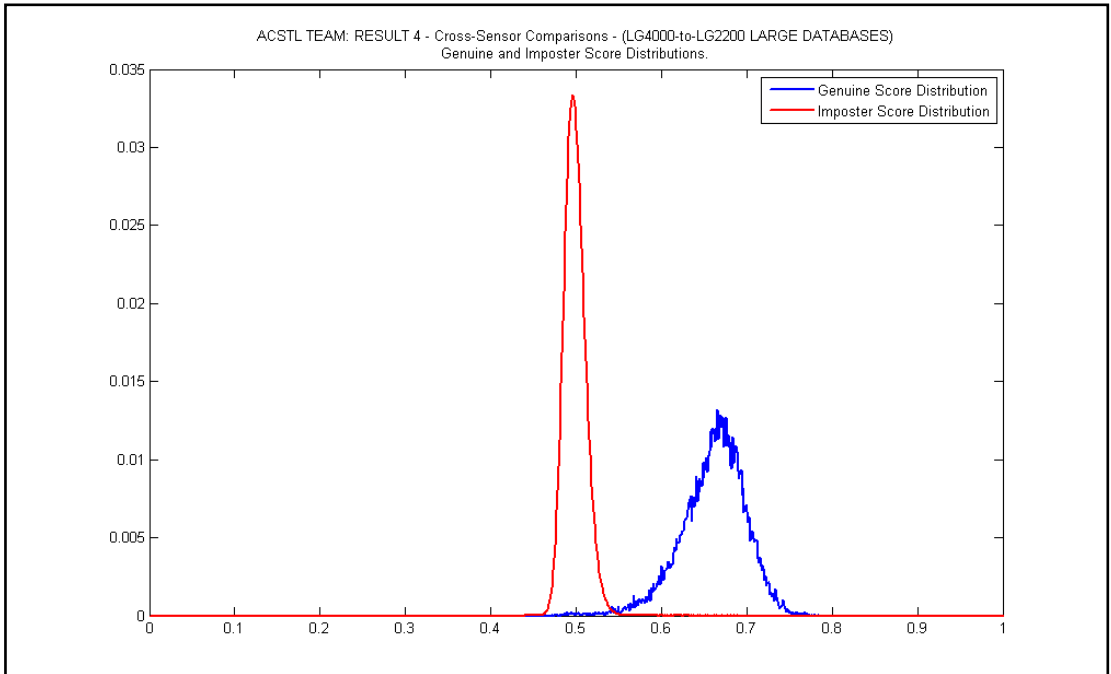

Fig.15: Imposter (μ=0.49882 σ=0.0124) and Genuine (μ=0.66065 σ=0.0380) Score Distributions (LinearY Scale)

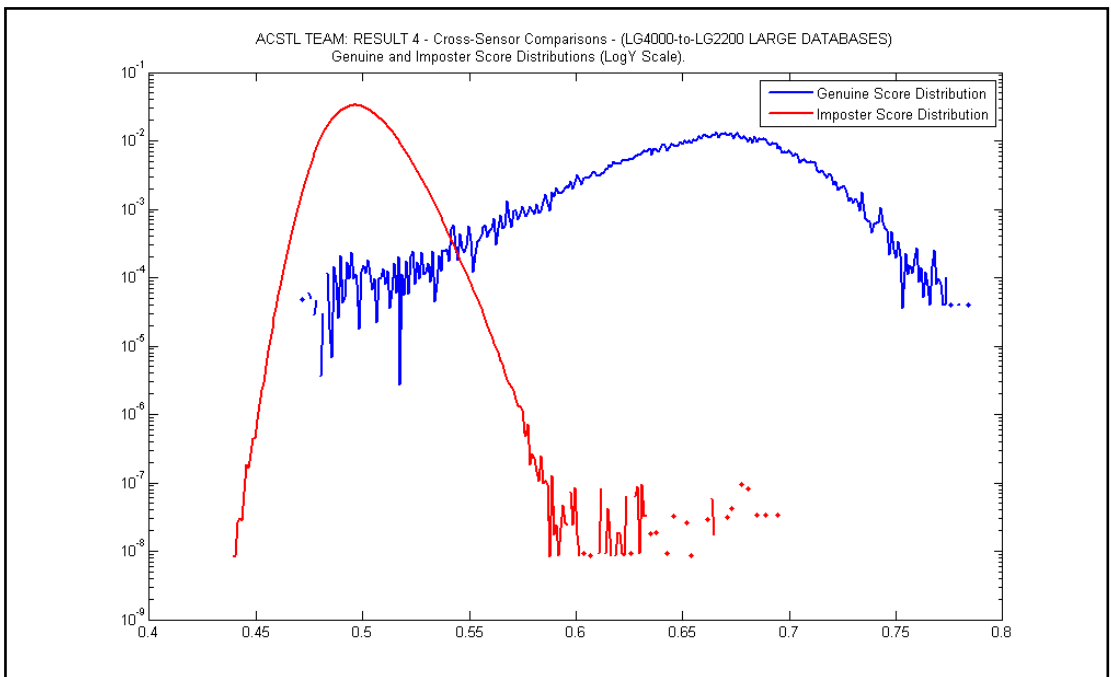

Fig.16: Imposter (μ=0.49882 σ=0.0124) and Genuine (μ=0.66065 σ=0.0380) Score Distributions (LogY Scale)

## IV. LG4000-TO-LG2200 IRIS RECOGNITION RESULTS OBTAINED ON LARGE LG4000 AND LG2200 DATABASES

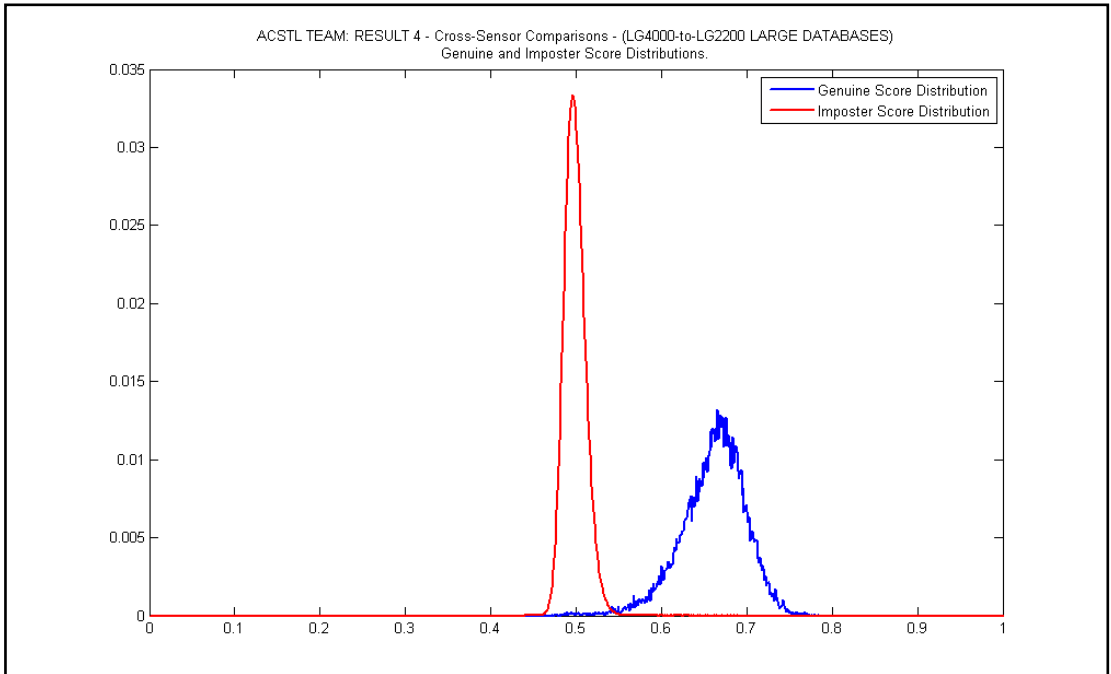

Fig.15: Imposter (μ=0.49882 σ=0.0124) and Genuine (μ=0.66065 σ=0.0380) Score Distributions (LinearY Scale)

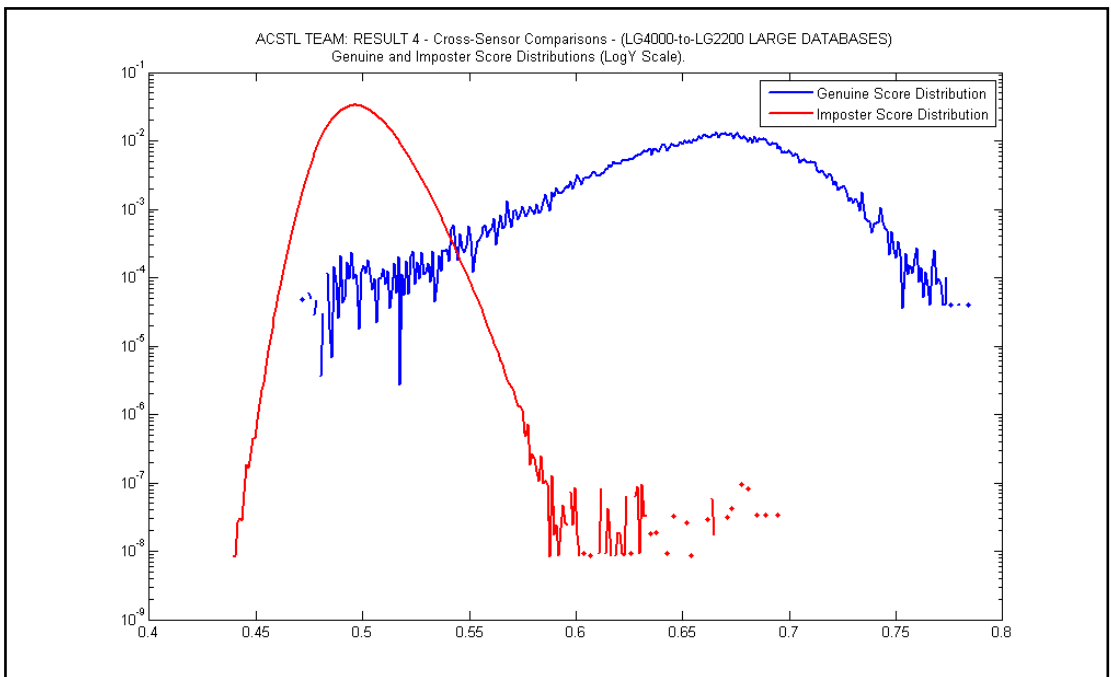

Fig.16: Imposter (μ=0.49882 σ=0.0124) and Genuine (μ=0.66065 σ=0.0380) Score Distributions (LogY Scale)

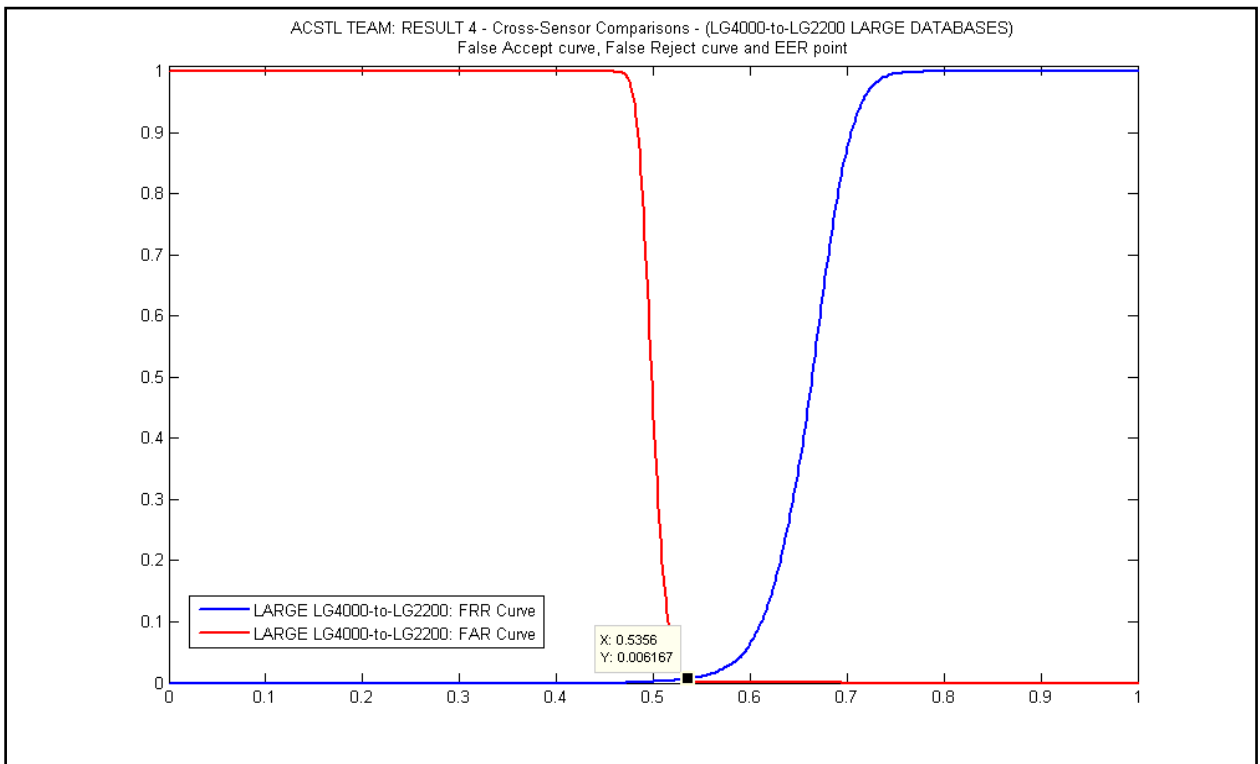

Fig.17: False Accept curve, False Reject curve, and EER point (LinearY Scale)

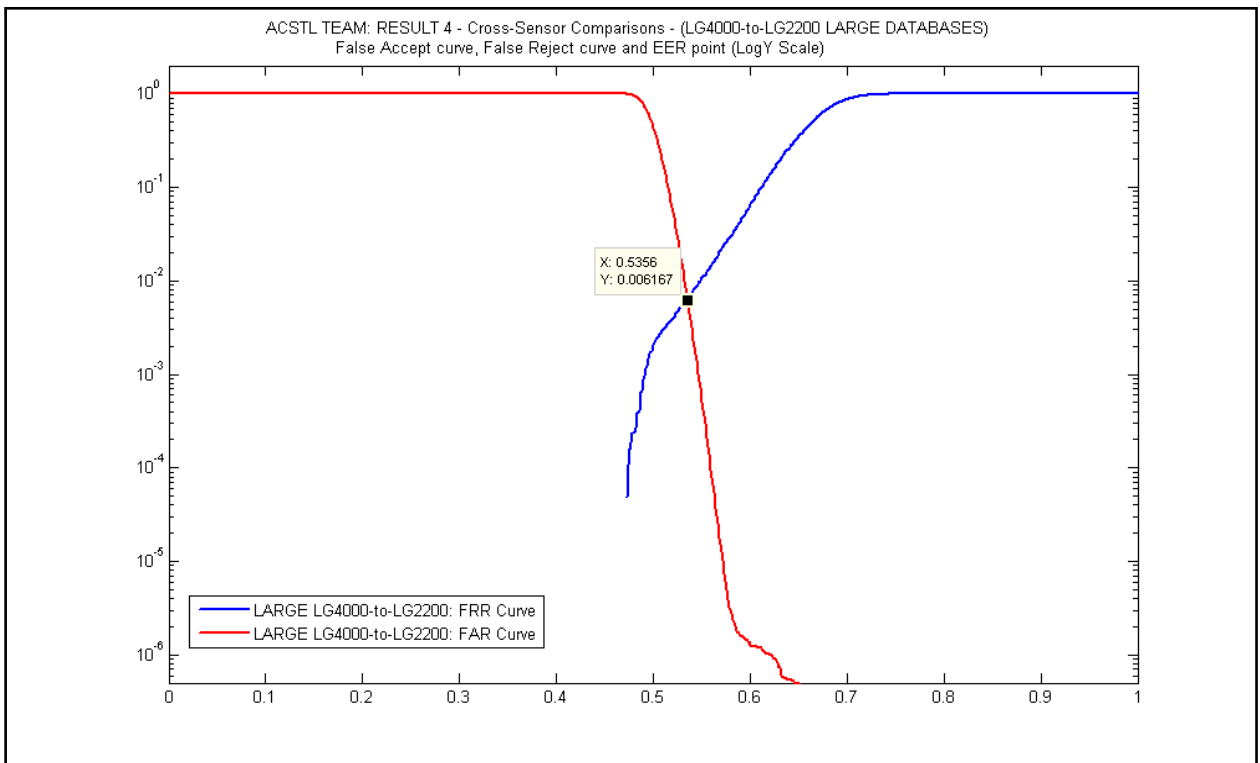

Fig.18: False Accept curve, False Reject curve, and EER point (LogY Scale)

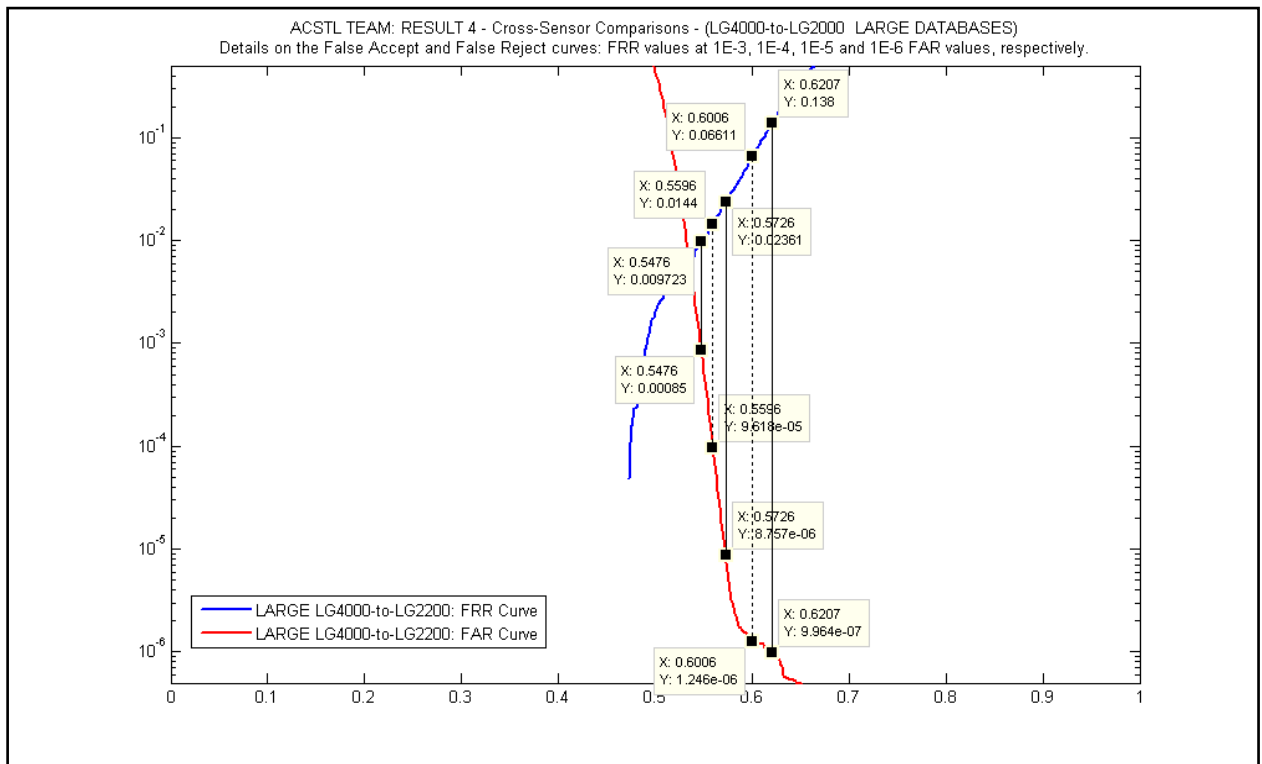

Fig.19: Details on the False Accept curve and False Reject curve (LogY Scale)

## V. ROC Curves

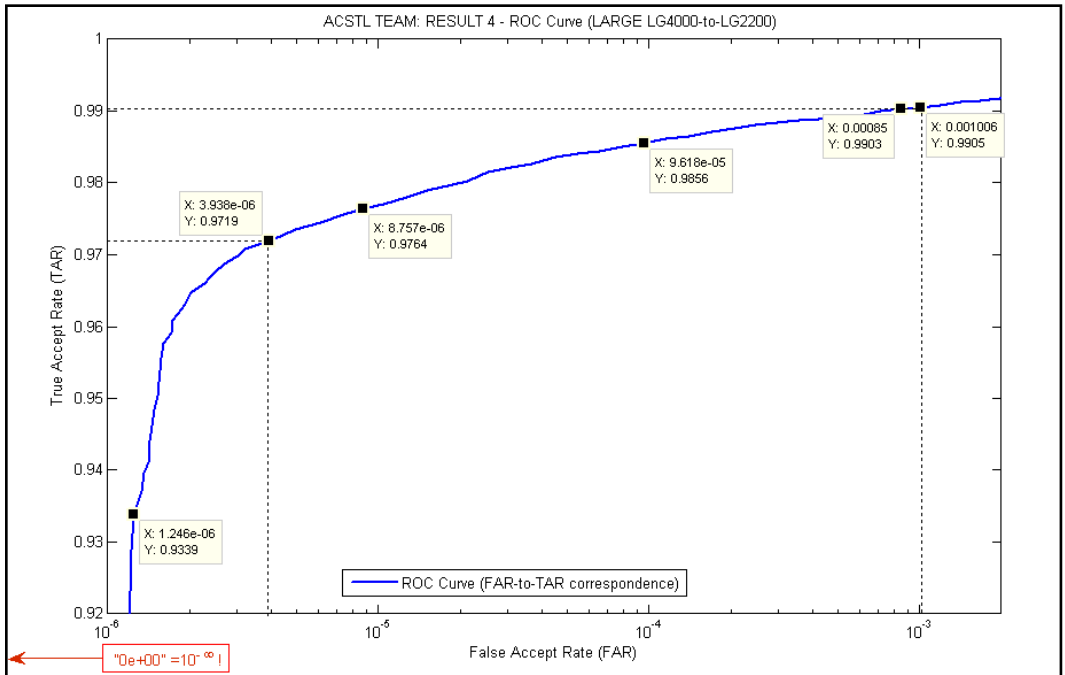

Fig.20: ROC Curve (FAR / TAR) for large LG4000-to-LG2200 comparison (our result – *ACSTL Cross -Sensor Comparison Competition Team 2013*, University of South-East Europe Lumina, RO)

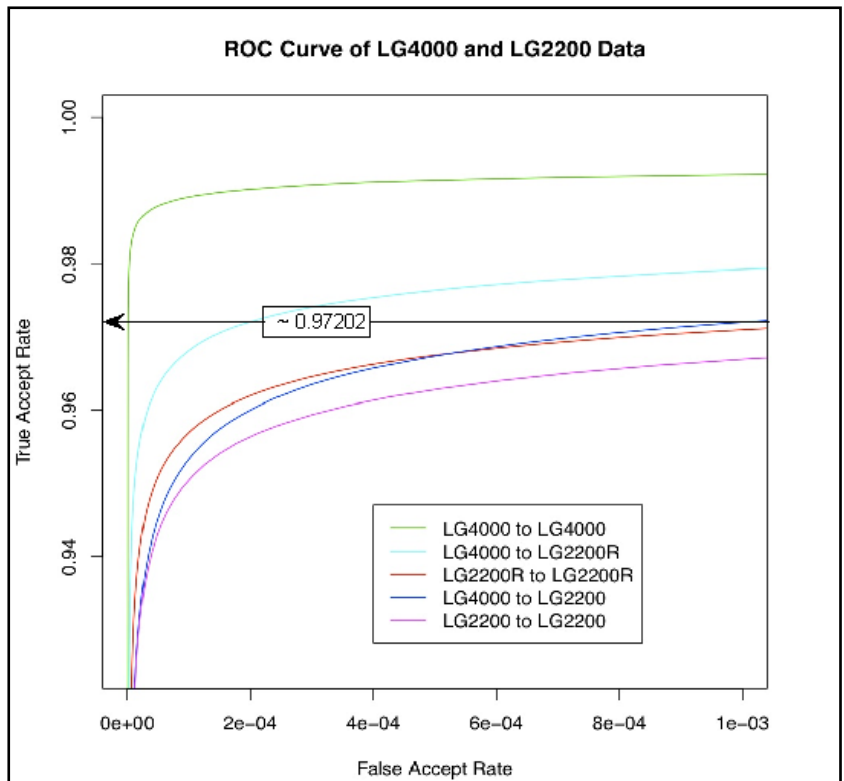

Fig.21: ROC Curve (FAR / TAR) for LARGE LG4000-to-LG2200 Comparison
(dark blue, the reference result – University of Notre Dame, USA)

## VI. *Cross-Sensor Comparison 2013* Database Related Issues

As proved by FAR, FRR and by the score distributions, multiple issues regarding the quality of the eye images can be identified in both LG2200 and LG4000 databases. The issues detected by us in the databases provided for the *Cross-Sensor Comparison Competition 2013* are the following:
- Cases of excessively dilated pupil;
- Cases of pupil covered by occlusion;
- Cases of patterned lenses;
- Cases of infrared-absorbing lenses;
- Cases of users enrolled with two IDs;
- Cases of iris templates enrolled under wrong user ID;
- Cases of iris textures strongly damaged by noise (especially blur and, sometimes, improper quantization);
- Some eye images are presented in non-axial gaze;
- Some eye images do not contain an iris image at all;

Images which presented one or more of these issues were verified both, automatically and visually. However, the images that could not be automatically detected as problematic were kept in the database.

Some of the issues listed above were also identified and mentioned in previous reports based on LG2200 databases, such as [3].

## VII. Comparison between the recognition rates obtained in this report and the reference results of *Cross-Sensor Comparison Competition 2013*

The reference results of *Cross-Sensor Comparison Competition 2013* are illustrated above, in Fig. 21. As it can be seen there, the reference ROC curve for LG4000-to-LG2200 comparison (the dark-blue line) starts its descent (right-side of the figure) approximately from the point (**1E-3**, **0.97202**) and ends in steep descent to approximately (**1E-4**, **0.93**).

The ROC curve corresponding to our result (Fig. 20) starts its descent (right-side of the figure) at (**1E-3**, **0.9905**), continues with a moderate descent rate through (**1E-4**, **0.9856**) and ends in steep descent to approximately (**1E-6**, **0.9339**). Hence, when compared to the reference ROC curve, our ROC curve reveals better biometric performances at both ends. We recall that, in order to preserve comparison compatibility with the reference cross-sensor iris recognition results, the mislabeled eye images were kept unchanged in the database, their influence being present in our results. We are pretty optimistic that even better recognition results could be and will be obtained by our team after the database clean-up.

As described above, our result has its chances to be better than the reference result, but the exact hierarchy could and should be established strictly by using the largest common set of images possible and an independent arbitrage. Unfortunately, these two needs were not anticipated in the 2013 edition of *Cross-Sensor Comparison Competition*. We are open to continue our work toward a decisive clear result of *Cross-Sensor Comparison Competition 2013* if these two conditions are fulfilled.

Another visual comparison can be made between our result and the one presented in Fig.9 from [4], where the best ROC curve starts its descent (right-side of the figure) approximately from the point (0.1, 0.99) and continues toward (0.01, 0.97). However, this ROC curve corresponds to LG4000-to-LG4000 comparison, hence it is at least twice weaker than those presented here in Fig. 20 and Fig. 21 from above.

## VIII. Conclusion

The intent of the Cross-Sensor Comparison Competition was "to analyze and improve the performance of iris recognition between two iris sensors from the same manufacturer" (LG). Cross-sensor comparison experimental results reported here shown that the procedure defined and simulated by *ACSTL Cross-Sensor Comparison Competition Team 2013* (USEEL) for migrating / upgrading LG2200 based to LG4000 based biometric systems leads to better LG4000-to-LG2200 cross-sensor iris recognition results in terms of user comfort (expressed as TAR – True Accept Rate) and system safety (expressed as FAR – False Accept Rate).

At the same level of security of 1E-3 FAR, our solution offers a higher level of user comfort (0.9905 TAR – our solution vs. 0.97202 TAR – the reference result). Also, at the same level of security of 1E-4 FAR, our solution offers a higher level of user comfort (0.9856 TAR – our solution vs. 0.93 TAR – the reference result). On the other hand, at the same level of user comfort of roughly 0.93 TAR, our solution allows a higher level of security (1E -6 FAR – our solution vs. 1E -4 FAR – the reference result). Of course, better results are expected after database clean-up.

## Acknowledgement

Thanks to Professors Lucian Grigore, Ragip Gokcel, Filip Stanciu, Constantin Balan, and Gheorghe Iubu, and thanks to Lumina Foundation and USEEL, the hardware cluster resources and the financial support necessary to secure our participation to *Cross-Sensor Comparison Competition 2013* were allotted in a timely manner.